\newif\ifElsev
\Elsevfalse

\ifElsev
\documentclass[preprint,authoryear]{elsarticle}
\else
\documentclass{article}
\fi
\usepackage{amsmath}
\usepackage{pstricks}
\usepackage{pst-node}
\usepackage{prooftree}
\usepackage{url}
\usepackage[all]{xy}
\usepackage{qsymbols}
\DeclareMathAlphabet{\mathpzc}{OT1}{pzc}{m}{it}

\ifElsev 
\else 
\usepackage{natbib}
\fi

\ifElsev 
\newproof{proof}{Proof} 
\else 
\newenvironment{proof}[1]{\begin{quotation}\noindent\textsf{Proof:} #1}%
{\(\Box\)\end{quotation}}
\fi
\newtheorem{prop}{Proposition}
\newcommand{\Coq}{\textsc{Coq}} 
\newcommand{\ie}{i.e.,~} 
\newcommand{\leut}{\ensuremath{\preceq}}

\newcommand{\lft}{\ensuremath{\ell}}
\newcommand{\rgt}{\ensuremath{\mathpzc{r}}}
\newcommand{\ltl}{\textit{``leads to a leaf''}}
\newcommand{\altl}{\textit{``always leads to a leaf''}}
\newcommand{\ob}{\mathfrak{o}}
\newcommand{\og}{\ensuremath{\ll\!}}
\newcommand{\fg}{\ensuremath{\hspace*{-1.5pt}\gg}}
\newcommand{\conva}{\ensuremath{\mathop{\vdash\! \raisebox{2pt}{\(\scriptstyle a\)}\! \dashv}}} 
\newcommand{\Alice}{\textsc{Alice}}
\newcommand{\Bob}{\textsc{Bob}}
\newcommand{\nat}{\ensuremath{\mathbb{N}}}

\usepackage{color} %
\definecolor{darkbrown}{cmyk}{.3,.75,.75,.15}
\definecolor{vertfonce}{rgb}{0,.5,0}

\newcommand{\rouge}[1]{{\color{red} #1}}

\definecolor{vertfonce}{rgb}{0,.5,0}

\usepackage{fancyhdr}
\newcommand{\coqdocleftpageheader}{\thepage\ -- \today}
\newcommand{\coqdocrightpageheader}{\today\ -- \thepage}
\def\_{\kern.08em\vbox{\hrule width.35em height.6pt}\kern.08em}

\newcommand{\coqdockw}[1]{\textsf{#1}}

\newcommand{\coqdocid}[1]{\textit{#1}}

\newcommand{\coqdoceol}{\setlength\parskip{0pt}\par}
\newcommand{\coqdocindent}[1]{\noindent\kern#1}

\newcommand{\ziginf}
{
  \begin{figure}[thb]
    \centering
    \begin{center}
\(\psframe[linewidth=2pt,framearc=.33](.4,-.3)(1.1,.3)
\)
  \begin{psmatrix}[rowsep=6pt,colsep=1.8pt] 
    &&&&[name=z]{$\bullet$}\\
   &&&[name=a]{$\bullet$} \\
      &&[name=b]{$\Box$} && [name=c]{$\bullet$} \\%
      &&&[name=d]{\textsf{zig}} && [name=e]{$\Box$}
      \ncline{a}{b} \ncline{a}{c} %
      \ncline{c}{d} \ncline{c}{e} %
      \ncline[linestyle=dotted]{z}{a}
\end{psmatrix}
\qquad
\raisebox{10pt}{{\huge $"=>"$}}
\qquad
  \(\psframe[linewidth=2pt,framearc=.33](0,-.3)(1.8,1.5)\)
\begin{psmatrix}[rowsep=6pt,colsep=1.8pt]
    &&&&[name=z]{$\bullet$}\\
    &&&[name=a]{$\bullet$} \\
      &&[name=b]{$\Box$} && [name=c]{$\bullet$} \\%
      &&&[name=d]{\textsf{zig}} && [name=e]{$\Box$}
      \ncline{a}{b} \ncline{a}{c} %
      \ncline{c}{d} \ncline{c}{e} %
      \ncline[linestyle=dotted]{z}{a}
\end{psmatrix} %
\end{center}

\begin{center}
  \rule{200pt}{1pt}
\end{center}

\begin{center}
  \(\psframe[linewidth=2pt,framearc=.33](-.1,-.2)(.7,.3)\)
  \textsf{zig}
\end{center}
    \caption{How cofix works on \textsf{zig} for \textit{is infinite}?}
    \label{fig:ziginf}
  \end{figure}
}
\newcommand{\figdollar}
{
\begin{figure}[tb]
  \centering
  \hspace*{-30pt} 
\(\begin{psmatrix}[colsep=20pt]
& [name=o]
  &{\ovalnode{a}{\Alice}} &{\ovalnode{b}{{\scriptstyle \Bob}}} & {\ovalnode{a1}{\Alice}} &{\ovalnode{b1}{{\scriptstyle \Bob}}}  & [name=c] & [name=d]
  &\\
  &[name=p]\phantom{\scriptstyle v+n, n} 
  &[name=e] {\scriptstyle v+n,n} 
  &[name=f] {\scriptstyle n+1,v+n} %
  &[name=e1] {\scriptstyle v+n+1,n+1} 
  &[name=f1] {\scriptstyle n+2,v+n+1} %
  &[name=h] \phantom{\scriptstyle 2n+1, 2n+2} 
  \ncarc[arrows=->,linestyle=dotted]{o}{p} %
  \ncarc[arrows=->,linestyle=dotted]{o}{a} %
  \ncarc[arrows=->]{a}{b} %
  \ncarc[arrows=->]{b}{a1} %
  \ncarc[arrows=->]{a1}{b1} %
  \ncarc[arrows=->,linestyle=dotted]{b1}{c} %
  \ncarc[arrows=->,linestyle=dotted]{c}{d} %
  \ncarc[arrows=->]{a}{e} %
  \ncarc[arrows=->]{b}{f} %
  \ncarc[arrows=->]{a1}{e1} %
  \ncarc[arrows=->]{b1}{f1} %
  \ncarc[arrows=->,linestyle=dotted]{c}{h} %
\end{psmatrix}
\)
  \caption{The \emph{dollar auction} game}
  \label{fig:dol_auct}
\end{figure}
}

\newcommand{\figfourstrat}
{\begin{figure}[thbp]
  \centering

    \hspace*{-30pt} 
\(\begin{psmatrix}[colsep=20pt]
  &{\ovalnode{a}{{\scriptstyle \Alice}}} &{\ovalnode{b}{{\scriptstyle \Bob}}} & {\ovalnode{a1}{{\scriptstyle \Alice}}} &{\ovalnode{b1}{{\scriptstyle \Bob}}}  & [name=c] & [name=d]
  &\\
  &[name=e] {\scriptstyle v+n,n} 
  &[name=f] {\scriptstyle n+1,v+n} %
  &[name=e1] {\scriptstyle v+n+1,n+1} 
  &[name=f1] {\scriptstyle n+2,v+n+1} %
  &[name=h] \phantom{\scriptstyle n+1,v+n} 
  \ncarc[arrows=->]{a}{b} %
  \ncarc[arrows=->]{b}{a1} %
  \ncarc[arrows=->]{a1}{b1} %
  \ncarc[arrows=->,linestyle=dotted]{b1}{c} %
  \ncarc[arrows=->,linestyle=dotted]{c}{d} %
  \ncarc[arrows=->,linewidth=.08]{a}{e} %
  \ncarc[arrows=->,linewidth=.08]{b}{f} %
  \ncarc[arrows=->,linewidth=.08]{a1}{e1} %
  \ncarc[arrows=->,linewidth=.08]{b1}{f1} %
  \ncarc[arrows=->,linewidth=.08,linestyle=dotted]{c}{h} %
\end{psmatrix}
\)

\bigskip

\textsf{dolAsBs}$_n$ aka \textbf{Always give up}

\bigskip

  \hspace*{-30pt} 
\(\begin{psmatrix}[colsep=20pt]
  &{\ovalnode{a}{{\scriptstyle \Alice}}} &{\ovalnode{b}{{\scriptstyle \Bob}}} & {\ovalnode{a1}{{\scriptstyle \Alice}}} &{\ovalnode{b1}{{\scriptstyle \Bob}}}  & [name=a2] & [name=b2]
  &\\
  &[name=e] {\scriptstyle v+n,n} 
  &[name=f] {\scriptstyle n+1,v+n} %
  &[name=e1] {\scriptstyle v+n+1,n+1} 
  &[name=f1] {\scriptstyle n+2,v+n+1} %
  &[name=h] \phantom{\scriptstyle n+1,v+n} 
  \ncarc[arrows=->]{a}{b} %
  \ncarc[arrows=->,linewidth=.08]{a}{e} %
  \ncarc[arrows=->,linewidth=.08]{b}{a1} %
  \ncarc[arrows=->]{b}{f} %
  \ncarc[arrows=->]{a1}{b1} %
  \ncarc[arrows=->,linewidth=.08]{a1}{e1} %
  \ncarc[arrows=->,linestyle=dotted,linewidth=.08]{b1}{a2} %
  \ncarc[arrows=->]{b1}{f1} %
  \ncarc[arrows=->,linestyle=dotted]{a2}{b2} %
  \ncarc[arrows=->,linestyle=dotted,linewidth=.08]{a2}{h} %
\end{psmatrix}
\)

\bigskip

\textsf{dolAsBc}$_n$ aka \textbf{\Alice{} abandons always and \Bob{} continues always}

\bigskip

  \hspace*{-30pt} 
\(\begin{psmatrix}[colsep=20pt]
  &{\ovalnode{a}{{\scriptstyle \Alice}}} &{\ovalnode{b}{{\scriptstyle \Bob}}} & {\ovalnode{a1}{{\scriptstyle \Alice}}} &{\ovalnode{b1}{{\scriptstyle \Bob}}}  & [name=a2] & [name=b2]
  &\\
  &[name=e] {\scriptstyle v+n,n} 
  &[name=f] {\scriptstyle n+1,v+n} %
  &[name=e1] {\scriptstyle v+n+1,n+1} 
  &[name=f1] {\scriptstyle n+2,v+n+1} %
  &[name=h] \phantom{\scriptstyle n+1,v+n} 
  \ncarc[arrows=->,linewidth=.08]{a}{b} %
  \ncarc[arrows=->,linewidth=.02]{a}{e} %
  \ncarc[arrows=->,linewidth=.02]{b}{a1} %
  \ncarc[arrows=->,linewidth=.08]{b}{f} %
  \ncarc[arrows=->,linewidth=.08]{a1}{b1} %
  \ncarc[arrows=->,linewidth=.02]{a1}{e1} %
  \ncarc[arrows=->,linestyle=dotted,linewidth=.02]{b1}{a2} %
  \ncarc[arrows=->,linewidth=.08]{b1}{f1} %
  \ncarc[arrows=->,linestyle=dotted,linewidth=.08]{a2}{b2} %
  \ncarc[arrows=->,linestyle=dotted,linewidth=.02]{a2}{h} %
\end{psmatrix}
\)

\bigskip

\textsf{dolAcBs}$_n$ aka \textbf{\Alice{} continues always and \Bob{} abandons always}

  \caption{Three strategy profiles}
  \label{fig:4_strat}
\end{figure}
}

\title{On the Rationality of Escalation}

\ifElsev \author[ens]{Pierre Lescanne\corref{cor1}}
\ead{Pierre.Lescanne@ens-lyon.fr}
\author[ens]{Matthieu Perrinel}
\address[ens]{University of Lyon, ENS de Lyon, CNRS (LIP), U. Claude Bernard de Lyon \\
 46 all\'ee d'Italie, 69364 Lyon, France}
\cortext[cor1]{Corresponding author: \textsf{Pierre.Lescanne@ens-lyon.fr}, {\scriptsize tel: +33 472728683, fax: +33 472728080}}
\else
\author{Pierre Lescanne and Matthieu Perrinel\\
University of Lyon, ENS de Lyon, \\CNRS (LIP), U. Claude Bernard de Lyon \\
 46 all\'ee d'Italie, 69364 Lyon, France}
\fi
\begin{document}

\ifElsev \else 
\maketitle 
\pagestyle{plain}
\thispagestyle{empty}
 \fi

\begin{abstract}
  Escalation is a typical feature of infinite games.  Therefore tools conceived for studying infinite mathematical structures, namely those deriving
  from \emph{coinduction} are essential.  Here we use coinduction, or backward coinduction (to show its connection with the same concept for finite
  games) to study carefully and formally the infinite games especially those called \emph{dollar auctions}, which are considered as the paradigm of
  escalation.  Unlike what is commonly admitted, we show that, provided one assumes that the other agent will always stop, bidding is rational,
  because it results in a subgame perfect equilibrium.  We show that this is not the only rational strategy profile (the only subgame perfect
  equilibrium).  Indeed if an agent stops and will stop at every step, we claim that he is rational as well, if one admits that his opponent will
  never stop, because this corresponds to a subgame perfect equilibrium.  Amazingly, in the infinite dollar auction game, the behavior in which both
  agents stop at each step is not a Nash equilibrium, hence is not a subgame perfect equilibrium, hence is not rational.  The right notion of
  rationality we obtain fits with common sense and experience and remove all feeling of paradox.

\ifElsev
\else
\fi
\end{abstract}

\ifElsev
\maketitle
\fi

\section{Introduction}

Escalation takes place in specific sequential games in which players continue although their payoff decreases on the whole.  The \emph{dollar auction}
game has been presented by \citet{Shubik:1971} as the paradigm of escalation.  He noted that, even though their cost (the opposite of the payoff)
basically increases, players may keep bidding.  This attitude is considered as inadequate and when talking about escalation, \citet{Shubik:1971} says
this is a paradox, \citet{oneill86:_inten_escal_and_dollar_auction} and \citet{leininger89:_escal_and_coop_in_confl_situat} consider the bidders as
irrational, \citet{gintis00:_game_theor_evolv} speaks of \emph{illogic conflict of escalation} and \citet{colman99:_game_theor_and_its_applic} calls
it \textit{Macbeth effect} after Shakespeare's play.  In contrast with these authors, in this paper, we prove using a reasoning conceived for
infinite structures that escalation is logic and that agents
are rational, therefore this is not a paradox and we are led to assert that Macbeth is somewhat rational.

This escalation phenomenon occurs in infinite sequential games and only there.  Therefore it must be studied with adequate tools, \ie in a framework
designed for mathematical infinite structures.  Like \citet{Shubik:1971} we limit ourselves to  two players only.  In auctions,
this consists in the two players bidding forever.  This statement is based on the common assumption that a player is rational if he adopts a strategy
which corresponds to a \emph{subgame perfect equilibrium}.  To characterize this equilibrium the above cited authors consider a finite restriction of
the game for which they compute the subgame perfect equilibrium by \emph{backward induction}\footnote{What is called ``backward induction'' in game
  theory is roughly what is called ``induction'' in logic.}.  In practice, they add a new hypothesis on the amount of money the bidders are ready to
pay, also called the limited bankroll.  In the amputated game, they conclude that there is a unique subgame perfect equilibrium.  This
consists in both agents giving up immediately, not starting the auction and adopting the same choice at each step.  In our formalization in infinite
games, we show that extending that case up to infinity is not a subgame perfect equilibrium and we found two subgame perfect
equilibria, namely the cases when one agent continues at each step and the other leaves at each step.  Those equilibria which correspond to rational attitudes account for the phenomenon of escalation.

The origin of the misconception that concludes the irrationality of escalation is the belief that properties of infinite mathematical objects can be
extrapolated from properties of finite objects.  This does not work.  As \citet{fagin93:_finit_model_theor_person_persp} recalls, ``most of the
classical theorems of logic [for infinite structures] fail for finite structures'' (see~\citet{EF-finite-mt} for a full development of the finite model
theory).  The reciprocal holds obviously ``most of the results which hold for finite structures, fail for infinite structures''.  This has been
beautifully evidenced in mathematics, when \citet{weierstrass72} has exhibited his function:
\begin{displaymath}
f(x)=\sum_{n=0}^\infty b^n\cos(a^n x \pi).
\end{displaymath}
Every finite sum is differentiable and the limit, \ie the \emph{infinite} sum, is not.  To give another
picture, infinite games are to finite games what fractal curves are to smooth curves \citep{edgar08:fract}.  In game theory the error done by the
ninetieth century mathematicians (Weierstrass quotes Cauchy, Dirichlet and Gauss) would lead to the same issue: wrong assumptions.  With what we are
concerned, a result that holds on finite games does not hold necessarily on infinite games and vice-versa.  More specifically equilibria on finite
games are not preserved at the limit on infinite games.  In particular, we cannot conclude that, whereas the only rational attitude in finite dollar
auction would be to stop immediately, it is irrational to escalate in the case of an infinite auction.  We have to keep in mind that in the case of
escalation, the game is infinite, therefore reasoning made for finite objects are inappropriate and tools specifically conceived for infinite objects
should be adopted.  Like Weierstrass' discovery led to the development of function series, logicians have invented methods for deductions on infinite structures and
the right framework for reasoning logically on infinite mathematical objects is called \emph{coinduction}.

Like induction, coinduction is based on a fixpoint, but whereas induction is based on the least fixpoint, coinduction is based on the greatest
fixpoint, for an ordering we are not going to describe here as it would go beyond the scope of this paper.  Attached to induction is the concept of
inductive definition, which characterizes objects like finite lists, finite trees, finite games, finite strategy profiles, etc.  Similarly attached to
coinduction is the concept of coinductive definition which characterizes streams (infinite lists), infinite trees, infinite games, infinite strategy
profiles etc.  An inductive definition yields the least set that satisfies the definition and a coinductive definition yields the greatest set that
satisfies the definition.  Associated with these definitions we have inference principles.  For induction there is the famous \emph{induction
  principle} used in backward induction.
On coinductively defined sets of objects there is a principle like induction principle which uses the fact that the set satisfies the definition
(proofs by case or by pattern) and that it is the largest set with this property.  Since coinductive definitions allow us building infinite objects,
one can imagine constructing a specific category of objects with ``loops'', like the infinite word $(abc)^{`w}$ (\ie $abcabcabc...$) which is made by
repeating the sequence $abc$ infinitely many times (other examples with trees are given in Section~\ref{sec:coind}, with infinite games and strategy
profiles in Section~\ref{sec:dol}).  Such an object is a fixpoint, this means that it contains an object like itself. For instance $(abc)^{`w} =
abc(abc)^{`w}$ contains itself. We say that such an object is defined as a cofixpoint.  To prove a property $P$ on a cofixpoint $\ob=f(\ob)$, one
assumes $P$ holds on $\ob$ (the $\ob$ in $f(\ob)$), considered as a sub-object of $\ob$.  If one can prove $P$ on the whole object (on $f(\ob)$), then
one has proved that $P$ holds on~$\ob$.  This is called the \emph{coinduction principle} a concept which comes from \citet{DBLP:conf/tcs/Park81},
\citet{DBLP:journals/tcs/MilnerT91}, and \cite{aczel88:_non_well_found_sets} (see also \citep{pair70:_concer_syntax_of_algol}) and was introduced in
the framework we are considering by \citet{DBLP:conf/types/Coquand93}.  \citet{DBLP:journals/toplas/Sangiorgi09} gives a good survey with a complete
historical account.  To be sure not be entangled, it is advisable to use a proof assistant that implements coinduction to build and check the proof,
but reasoning with coinduction is sometimes so counter-intuitive that the use of a proof assistant is not only advisable but compulsory.  For
instance, we were, at first, convinced that the strategy profile consisting in both agents stopping at every step was a Nash equilibrium, like in the
finite case, and only failing in proving it mechanically convinced us of the contrary and we were able to prove the opposite.  In our case we have
checked every statement using \Coq{} and in what follows a sentence like ``we have prover that ...''  means that we have succeeded in building a
formal proof in \Coq.

\subsection*{Backward coinduction as a method for proving invariants}
\label{sec:backw-coinduct-as}

In infinite strategy profiles, the coinduction principles can be seen as follows:  a~property which holds on a strategy profile of  an infinite
extensive game is an \emph{invariant}, \ie a
property that stays always true, along the temporal line and to prove that this is an invariant one proceeds back to the past.  Therefore the name
\emph{backward coinduction} is appropriate, since it proceeds backward the time, from future to past.

\subsection*{Backward induction vs backward coinduction}
\label{sec:backw-induct-vs}

One may wonder the difference between the classical method, which we call \emph{backward induction} and the new method we propose, which we call
\emph{backward coinduction}.  The main difference is that backward induction starts the reasoning from the leaves, works only on finite games and does
not work on infinite games (or on finite strategy profiles), because it requires a well-foundedness to work properly, whereas \emph{backward
  coinduction} works on infinite games (or on infinite strategy profiles).  Coinduction is unavoidable on infinite games, since the methods that
consists in ``cutting the tail'' to get a finite game or a finite strategy profile cannot solve the problem or even approximate it. Using backward induction
to a game which is intrinsically infinite like the escalation in the dollar auction was a mistake.
It is indeed the
same erroneous reasoning as this of the predecessors of Weierstrass who concluded that since:
\begin{displaymath}
`A p`:\nat,  f(x)=\sum_{n=0}^p b^n\cos(a^n x \pi),
\end{displaymath}
is differentiable everywhere then
\begin{displaymath}
f(x)=\sum_{n=0}^\infty b^n\cos(a^n x \pi).
\end{displaymath}
is differentiable everywhere whereas it is differentiable nowhere. 

 Much earlier, during the IV$^{th}$ century BC, the improper use of inductive reasoning allows Parmenides and Zeno to negate
motion and leads to \emph{Zeno's paradox of Achilles and the tortoise}.  This paradox was reported by Aristotle as follows:
\begin{quotation}
 \emph{``In a race, the quickest runner can never overtake the slowest, since the pursuer must first reach the point whence the pursued started, so
   that the slower must always hold a lead.''}

\hfill\emph{Aristotle}, Physics VI:9, 239b15
\end{quotation}
Zeno's reasoning is correct, because by induction, one can prove that Achilles will never overtake
the tortoise, but we know by experience that this is not the case, hence the paradox.  Zeno error was to apply induction to an infinite object, he
should have used coinduction if he would have known this concept.

\subsection*{Von Neumann and coinduction}
\label{sec:von-neumann}

As one knows, von Neumann \citep{neumann28:_zur_theor_gesel,neumann44:_theor_games_econom_behav} is the creator of game theory, whereas extensive
games and equilibrium in non cooperative games are due
to \cite{Kuhn:ExtGamesInfo53} and \cite{Nash50}.  In the spirit of their creators all those games are finite and backward induction is the basic principle
 for computing subgame perfect equilibria \citep{selten65:_spiel_behan_eines_oligop_mit}.  This is not surprising since
\cite{neumann25:_axiom_mengen} is also at the origin of the role of well-foundedness in set theory despite he left a door open for a not well-founded
membership relation.  As explained by
\cite{DBLP:journals/toplas/Sangiorgi09}, research on anti-foundation initiated by \cite{mirimanoff17:_les_antim_de_russel_et} are at the origin
of coinduction and were not well known  until the work of \cite{aczel88:_non_well_found_sets}.

\subsection*{Why in infinite plays, agents do not have a utility ?}
\label{sec:why-infinite-plays}

In our framework, in an infinite play (a play that runs forever, \ie that does not lead to a leaf) no agent has a utility.  People might say that this an anomaly, but we
claim that this is perfectly sensible.  Let us affirm that in arbitrary long plays, which lead to a leaf, all agents have a utility.  Only in plays
that diverge, it is the case that agents have no utility.  This fits well with \cite{binmore88:_model_ration_player_part_ii} statements \emph{``The use of
  computing machines (automata) to model players in an \emph{evolutive} context is presumably uncontroversial ... machines are also appropriate for
  modeling players in an \emph{eductive} context''}.  Here we are concerned by the eductive context where \emph{``equilibrium is achieved through
  careful reasoning by the agents before and during the play of the game''} \cite[loc. cit]{binmore88:_model_ration_player_part_ii}.  By automaton, we
mean any model of computation\footnote{Our model of computation is this of the calculus of inductive construction, a kind of $`l$-calculus behind
  \Coq{} \citep{turing37:_comput_and_lambda_defin}.}, since all the models of computation are equivalent by Church thesis.  If an agent is modeled
by an automaton, this means also that the function that computes the utility for this agent is also modeled by an automaton.  It seems then sensible
that one cannot compute the utility of an agent for an infinite play, since computing is a finite process working on finite data (or at least data
that are finitely described).  Since the agent cannot compute the utility of an infinite play, no sensible value can be attributed to him.  If one wants absolutely to
assign a value to an infinite play, one must abandon the automaton framework and moreover this value should be the limit of a sequence of values,
which does not exist in most of the cases.\footnote{If utilities are natural numbers, it exists only if the sequence is stationary, which is not the
  case in escalation.}  For instance, in the case of the dollar auction
(Section~\ref{sec:dol}), the  utility associated with the unique infinite play are the sequence $..., v+n, n+1, v+n+1, n+2, ...$
Therefore considering that in infinite plays, agents have no utility is perfectly consistent with a modeling of agents by automata.  
By the way, does an agent care about a payoff he (she) receives in infinitely many years?  Will he (she) adapt his (her) strategy on this?

\subsection*{Proof assistants vs automated theorem provers}
\label{sec:proof-assistant-vs}

\Coq{} is a proof assistant built by \emph{\citet{Coq:manual}}, see~\citet{BertotCasterant04} for a good introduction and notice that they call it ``interactive
theorem provers'', which is a strict synonymous.  Despite both deal with theorems and their proofs and are mechanized using a computer, proof assistants are not automated theorem
provers. In particular, they are much more expressive than automated theorem provers and this the reason why they are interactive.  For instance,
there is no automated theorem prover implementing coinduction.  Proof assistants are automated only for elementary steps and interactive for the rest.
A specificity of a proof assistant is that it builds a mathematical object called a (formal) proof which can be checked independently, copied, stored and
exchanged.   Following \cite{harrison-notices} and \cite{dowek07:_les_metam_du_calcul}, we can consider that they are the tools of the mathematicians
of the XXI${^\textrm{th}}$ century.  Therefore using a proof assistant is a highly mathematical modern activity.

The mathematical development presented here corresponds to a \Coq{} script\footnote{A \emph{script} is a list of commands of a proof assistant.} which can be found on the following url's:

\centerline{\url{http://perso.ens-lyon.fr/pierre.lescanne/COQ/EscRat/}}

\centerline{\url{http://perso.ens-lyon.fr/pierre.lescanne/COQ/EscRat/SCRIPTS/}}

\medskip

\subsection*{Structure of the paper}
\label{sec:structure-paper}

The paper is structured as follows. In Section~\ref{sec:coind} we present coinduction illustrated by the example of infinite binary trees.  In Section
\ref{sec:fin-inf-games} we present infinite games.
In Section~\ref{sec:inf_strat}, we
introduce the core concept of infinite strategy profile which allows us presenting equilibria in Section~\ref{sec:equi}.  The dollar auction game is presented in
Section~\ref{sec:dol} and the escalation is discussed in Section~\ref{sec:esc}.  Readers who want to have a quick idea about the results of this paper
on the rationality of
escalation are advised to read sections~\ref{sec:dol}, \ref{sec:esc} and \ref{sec:concl}.

\subsection*{Related works}

To our knowledge, the only application of
coinduction to extensive game theory has been made by \cite{capretta:2007} who uses coinduction to define only
common knowledge not equilibria in infinite games.  Another strongly connected work is this of \citet{coupet-grimal03:_axiom_of_linear_temp_logic} on temporal logic.
  Other applications are on representation of real numbers by infinite sequences
\citep{bertot07:_affin_funct_and_series_with,DBLP:conf/flops/Julien08} and implementation of streams (infinite lists) in electronic circuits
\citep{DBLP:journals/fac/Coupet-GrimalJ04}.  An ancestor of our description of infinite games and infinite strategy profiles is the constructive
description of finite games, finite strategy profiles, and equilibria by \citet{vestergaard06:IPL}.   
\citet{DBLP:journals/corr/abs-0904-3528} introduces the framework of infinite games with more detail.  Infinite games are introduced in
\citet{osborne94:_cours_game_theory} and \citet{osborne04a} using histories, but this is not algorithmic and therefore not amenable to formal proofs
and coinduction.  

Many authors have studied infinite games (see for instance \citet{martin98:_deter_of_black_game,DBLP:conf/dagstuhl/Mazala01}),
but except the name ``game'' (an overloaded one), those games have nothing to see with infinite extensive games as presented in this paper.  The
infiniteness of Blackwell games for instance is derived from a topology, by adding real numbers and probability. 
\cite{DBLP:journals/toplas/Sangiorgi09} mentioned the connection between Ehrenfeucht-Fra{\"i}ss{\'e} games \citep{EF-finite-mt} and coinduction, but
the connection with extensive games is extremely remote.

\section[coinduction, infinite trees]{Coinduction and infinite binary trees}
\label{sec:coind}

As an example of a coinductive definition consider this of \emph{lazy binary trees}, \ie finite and infinite binary
trees.

\begin{figure}[thb]
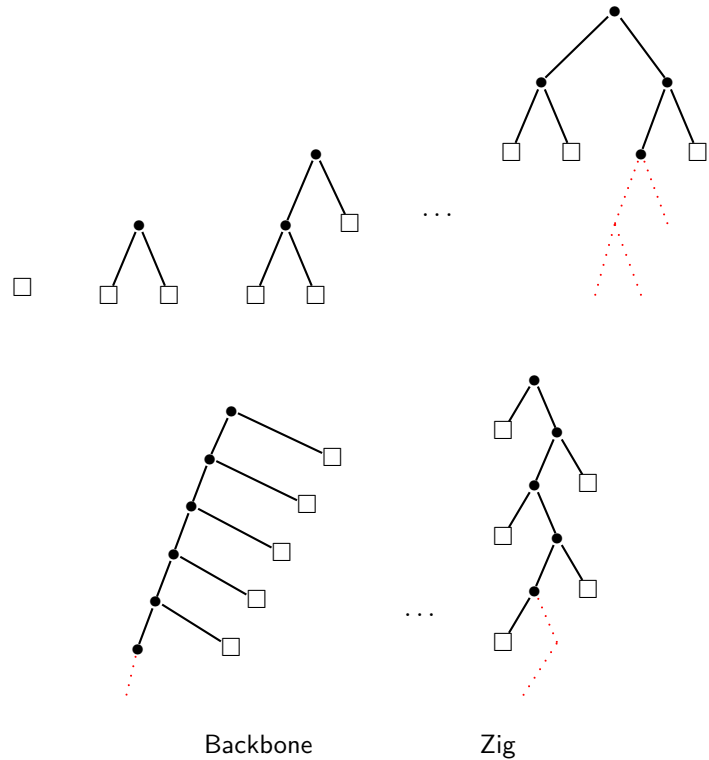

  \begin{center}
    $\Box$\qquad
    \begin{psmatrix}[rowsep=15pt,colsep=5pt]
      &&[name=a]{$\bullet$} & \\
      &[name=b]{$\Box$} && [name=c]{$\Box$} %
      \ncline{a}{b} \ncline{a}{c}
    \end{psmatrix}\qquad
    \begin{psmatrix}[rowsep=15pt,colsep=5pt]
      &&&[name=a]{$\bullet$} & \\
      &&[name=b]{$\bullet$} && [name=c]{$\Box$} \\%
      &[name=d]{$\Box$}&&[name=e]{$\Box$} 
      \ncline{a}{b} \ncline{a}{c} \ncline{b}{d} \ncline{b}{e}
    \end{psmatrix}
    \qquad \raisebox{30pt}{$\ldots$}
    \begin{psmatrix}[rowsep=15pt,colsep=5pt]
      &&&&&&&[name=a]{$\bullet$} \\
      &&&&[name=b]{$\bullet$} &&&&& [name=c]{$\bullet$} \\%
      &&&[name=d]{$\Box$}&&[name=e]{$\Box$}
      &&&[name=f]{$\bullet$}&&[name=g]{$\Box$} \\
      &&&&&&&[name=h]&&[name=i] \\ %
      &&&&&&[name=j]&&[name=k] %
      \ncline{a}{b} \ncline{a}{c} %
      \ncline{b}{d} \ncline{b}{e} \ncline{c}{f} \ncline{c}{g} %
      \rouge{\ncline[linestyle=dotted,linecolor=red]{f}{h}} \ncline[linestyle=dotted,linecolor=red]{f}{i}
      \rouge{\ncline[linestyle=dotted,linecolor=red]{h}{j}} \ncline[linestyle=dotted,linecolor=red]{h}{k}
    \end{psmatrix}
  \end{center}

  \medskip

  \begin{center}
    \begin{psmatrix}[rowsep=6pt,colsep=1.8pt]
      &&&&&&&[name=a]{$\bullet$} \\
      &&&&&&[name=b]{$\bullet$} &&&&& [name=c]{$\Box$} \\%
      &&&&&[name=d]{$\bullet$}&&&&&[name=e]{$\Box$} \\
      &&&&[name=f]{$\bullet$}&&&&&[name=g]{$\Box$} \\
      &&&[name=h]{$\bullet$}&&&&&[name=i]{$\Box$} \\ %
      &&[name=j]{$\bullet$}&&&&&[name=k]{$\Box$}\\ %
      &[name=l] %
      \ncline{a}{b} \ncline{a}{c} %
      \ncline{b}{d} \ncline{b}{e} %
      \ncline{d}{f} \ncline{d}{g} %
      \ncline{f}{h} \ncline{f}{i} %
      \ncline{h}{j} \ncline{h}{k} %
      \ncline[linestyle=dotted,linecolor=red]{j}{l}
    \end{psmatrix}
    \qquad\raisebox{30pt}{$\ldots$}\qquad
    \begin{psmatrix}[rowsep=8pt,colsep=1.8pt]
      &&&&[name=a]{$\bullet$} \\
      &[name=b]{$\Box$} &&&&& [name=c]{$\bullet$} \\%
      &&&&[name=d]{$\bullet$}&&&&&[name=e]{$\Box$} \\
      &[name=f]{$\Box$}&&&&&[name=g]{$\bullet$} \\
      &&&&[name=h]{$\bullet$}&&&&&[name=i]{$\Box$} \\ %
      &[name=j]{$\Box$}&&&&&[name=k]\\ %
      &&&[name=l] \ncline{a}{b} \ncline{a}{c} %
      \ncline{c}{d} \ncline{c}{e} \ncline{d}{f} \ncline{d}{g} \ncline{g}{h} \ncline{g}{i} \ncline{h}{j} \ncline[linestyle=dotted,linecolor=red]{h}{k}
      \ncline[linestyle=dotted,linecolor=red]{k}{l}
    \end{psmatrix}
  \end{center}

  \begin{center}
    \textsf{Backbone} \qquad \qquad \qquad \textsf{Zig}
  \end{center}
  \caption{Coinductive binary trees}
  \label{fig:lazy}
\end{figure}
\medskip

\begin{quotation}
  A \textsf{coinductive} \emph{binary tree} (or a lazy binary tree or a finite-infinite binary tree) is
 \ifElsev \begin{itemize}[$\star$] \else \begin{itemize} \fi
  \item either the empty binary tree $\Box$,
  \item or a binary tree of the form $t `. t'$, where $t$ and $t'$ are binary trees.
  \end{itemize}
\end{quotation}

By the keyword \textsf{coinductive} we mean that we define a coinductive set of objects, hence we accept infinite objects.  Some coinductive binary trees are
given on Fig.~\ref{fig:lazy}.  We define on a coinductive binary tree a predicate which has also a coinductive definition:

\begin{quotation}
  A binary tree is \emph{infinite} if (coinductively)
\ifElsev \begin{itemize}[$\star$] \else \begin{itemize} \fi 
  \item either its left subtree is \emph{infinite}
  \item or its right subtree is \emph{infinite}.
  \end{itemize}
\end{quotation}

We define two trees that we call \textsf{zig} and \textsf{zag}.
\begin{quotation}
  \textsf{zig} and \textsf{zag} are defined together as \emph{cofixpoint}s as follows:
\ifElsev \begin{itemize}[$\star$] \else \begin{itemize} \fi
  \item \textsf{zig} has  $\Box$ as left subtree and \textsf{zag} as right subtree,
  \item \textsf{zag} has \textsf{zig} as left subtree and $\Box$  as right subtree.
  \end{itemize}
\end{quotation}
This says that \textsf{zig} and \textsf{zag} are the greatest solutions\footnote{In this case, the least solutions are uninteresting as they are
  objects nowhere defined. Indeed there is no basic case in the inductive definition.} of the two simultaneous equations:
\begin{eqnarray*}
  \mathsf{zig} &=& \Box  `.  \mathsf{zag}\\
  \mathsf{zag} &=& \mathsf{zig}  `. \Box
\end{eqnarray*}

\ziginf{}

It is common sense that \textsf{zig} and \textsf{zag} are infinite, but to prove that \emph{``\textsf{zig} is infinite''} using the \textsf{cofix}
tactic\footnote{The \textsf{cofix} tactic is a method proposed by the proof assistant \Coq{} which implements coinduction on cofixpoint objects.
  Roughly speaking, it attempts to prove that a property is an \emph{invariant}, by proving it is preserved along the infinite object.  Here \emph{``
    is infinite'' } is such an invariant on \textsf{zig}.}, we do as follows: assume \emph{``\textsf{zig} is infinite''}, then \textsf{zag} is
infinite, from which we get that \emph{``\textsf{zig} is infinite''}.  Since we use the assumption on a strict subtree of \textsf{zig} (the direct
subtree of \textsf{zag}, which is itself a direct subtree of \textsf{zig}) we can conclude that the \textsf{cofix} tactic has been used properly and
that the property holds, namely that \emph{``\textsf{zig} is infinite''}.  This is pictured on Fig.\ref{fig:ziginf}, where the square box represents the predicate \emph{is infinite}.  Above the rule, there is the step of coinduction and below the rule the conclusion, namely that the whole \textsf{zig} is infinite.
We let the reader prove that \textsf{backbone} is infinite, where
\textsf{backbone} is the greatest fixpoint of the equation:
\begin{center}
  \textsf{backbone} \ = \ \textsf{backbone}  $`.\ \Box$
\end{center}
and is an infinite tree that looks like the skeleton of a  infinite centipede game as shown on Fig.\ref{fig:lazy} (see Section~\ref{sec:anot-exampl-infin}).

Interested readers may have a look
at~\citet*{coupet-grimal03:_axiom_of_linear_temp_logic,DBLP:journals/fac/Coupet-GrimalJ04,DBLP:journals/corr/abs-0904-3528,bertot05:_filter_coind_stream_applic_to_eratos_sieve,bertot07:_affin_funct_and_series_with}
  and especially \citet[chap.~13]{BertotCasterant04} for other examples of \textsf{cofix} reasoning.

\section{Finite and infinite games}
\label{sec:fin-inf-games}

As an intermediary between histories and strategy profiles, let us define finite and infinite games.  Traditionally, games are defined through trees
associated with utility function at the leaves.  Another approach which \cite{osborne04a} attributes to Rubinstein uses histories.  A third
approach proposed by \cite{vestergaard06:IPL} which fits well with inductive reasoning is to give an inductive definition of games.   To handle infinite games
we propose a coinductive definition. 
\begin{quotation}
  The type of \emph{Game}s is defined as a \textbf{coinductive} as follows:
  \begin{itemize}
  \item a \emph{Utility function} makes a \emph{Game},
  \item an \emph{Agent} and two \emph{Game}s make a \emph{Game}.
  \end{itemize}
\end{quotation}

A \emph{Game} is either a leaf (a terminal node) or a composed game made of an agent (the agent who has the turn) and two subgames (the formal definition in the \Coq{} vernacular
is given in the appendix~\ref{sec:coq_vern}).  We use the expression \coqdocid{gLeaf} $f$ to denote the leaf game associated with the utility function $f$ and the expression
\coqdocid{gNode} $a$ $g_l$ $g_r$ to denote the game with agent $a$ at the root and two subgames $g_l$ and $g_r$.

Hence one builds a finite game in two ways: either
a given utility function $f$ is encapsulated by the operator \coqdocid{gLeaf} to make the game (\coqdocid{gLeaf} \coqdocid{f}), or an agent \coqdocid{a} and two games $g_l$ and
$g_r$ are given to make the game (\coqdocid{gNode} \coqdocid{a} $g_l$ $g_r$).  Notice that in such games, it can be the case that the same
agent \coqdocid{a} has the turn twice in a row, like in the game (\coqdocid{gNode} \coqdocid{a} (\coqdocid{gNode} \coqdocid{a} $g_1$ $g_2$) $g_3$).

Concerning comparisons of utilities we consider a very general setting where a utility is no more that a type (a ``set'') with a preference which is a
preorder, i.~e., a transitive and reflexive relation, and which we write \leut. A preorder is enough for what we want to prove.
We assign to the leaves, a utility function which associates a utility to each agent.

We can also tell how we associate a history with a game or a history and a utility function with a game (see the \Coq{} script).  We will see in the next section how to associate a
utility with an agent in a game, this is done in the frame of a strategy profile, which is described now.

\section{Finite or infinite strategy profiles}
\label{sec:inf_strat}

In this section we define \emph{finite or infinite binary strategy profiles} or \emph{StratProf}s in short.  They are based on games which are
extensive (or sequential) games and in which each agent has two choices: {\lft} (left) and \rgt{} (right).\footnote{In pictures, we take a
  subjective point of view: \emph{left} and \emph{ right} are from the perspective of the agent.} In addition these games are infinite, we should say
``can be infinite'', as we consider both finite and infinite games.  We do not give explicitly the definition of a finite or infinite extensive game
since we do not use it in what follows, but it can be easily obtained by removing the choices from a strategy profile.  To define finite or
infinite strategy profiles, we suppose given a \emph{utility} and a \emph{utility function}.  As said, we define directly strategy profiles as they are the
only concept we are interested in.  Indeed an equilibrium is a strategy profile.

\begin{quotation}
  The type of \emph{StratProf}s is defined as a \textbf{coinductive} as follows:
\ifElsev \begin{itemize}[$\star$] \else \begin{itemize} \fi
  \item a \emph{Utility function} makes a \emph{StratProf}.
  \item an \emph{Agent}, a \emph{Choice} and two \emph{StratProf}s make a \emph{StratProf}.
  \end{itemize}
\end{quotation}

Basically\footnote{The formal definition in the \Coq{} vernacular is given in appendix~\ref{sec:coq_vern}.} an infinite strategy profile which is not
a \emph{leaf} is a \emph{node} with four items: an agent, a choice, two infinite strategy profiles.  A strategy profile is a game plus a choice at
each node.  Strategy profiles of the first kind are written $\og f \fg$ and strategy profiles of the second kind are written $\og a,c,s_l,s_r\fg$.  In
other words, if between the ``$\og$'' and the ``$\fg$'' there is one component, this component is a utility function and the result is a leaf strategy
profile and if there are four components, this is a compound strategy profile.  In what follows, we say that $s_l$ and $s_r$ are strategy subprofiles
of $\og a,c,s_l,s_r\fg$.  For instance, here are the drawing of two strategy profiles ($s_0$ and $s_1$):

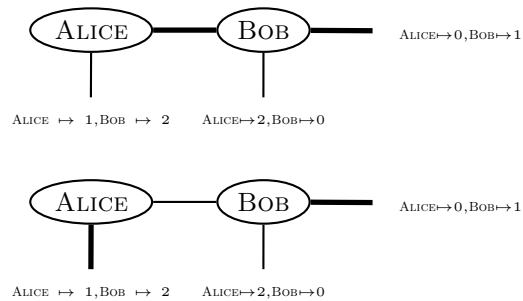
\[\begin{psmatrix}[colsep=3pt,rowsep=10pt]
  & {\ovalnode{a}{\Alice}} &&&&{\ovalnode{b}{\Bob}} &&&&&& [name=e]{\quad \scriptscriptstyle\Alice \mapsto 0, \Bob \mapsto 1}\\
  & [name=c]{\raisebox{-10pt}{$\scriptscriptstyle \Alice ~\mapsto~ 1, \Bob ~\mapsto~ 2$}} 
  &&&& [name=d]{\raisebox{-10pt}{$\scriptscriptstyle \Alice \mapsto 2, \Bob \mapsto 0$}} 
  \ncline[linewidth=.07]{a}{b}
  \ncline{a}{c}
  \ncline{b}{d}
  \ncline[linewidth=.07]{b}{e}
\end{psmatrix} %
\]

\bigskip

\[
\begin{psmatrix}[colsep=3pt,rowsep=10pt]
  &{\ovalnode{a}{\Alice}} &&&&{\ovalnode{b}{\Bob}} &&&&&& [name=e]{\quad \scriptscriptstyle\Alice \mapsto 0, \Bob \mapsto 1}  \\
  & [name=c]{\raisebox{-10pt}{$\scriptscriptstyle \Alice ~\mapsto~ 1, \Bob ~\mapsto~ 2$}}
  &&&&[name=d]{\raisebox{-10pt}{$\scriptscriptstyle \Alice \mapsto 2, \Bob \mapsto 0$}} 
  \ncline{a}{b}
  \ncline[linewidth=.07]{a}{c}
  \ncline{b}{d}
  \ncline[linewidth=.07]{b}{e}
\end{psmatrix}
\]

\medskip
\noindent which correspond to the expressions

  $s_0 =  \og$\emph{\Alice},\lft,$\og$\emph{\Bob}, \lft, $\og${\scriptsize $\Alice \mapsto 0, \Bob \mapsto 1$}$\fg$, %
                                                 $\og${\scriptsize$\Alice \mapsto 2, \Bob \mapsto 0$}$\fg$ $\fg$,\\
  \hspace*{78pt} $\og$ $\scriptstyle \Alice ~\mapsto~ 1, \Bob ~\mapsto~ 2$ $\fg$ $\fg$

and

  $s_1 = \og${\Alice},\rgt{},$\og$\emph{\Bob},\lft,$\og${\scriptsize $\Alice \mapsto 0, \Bob \mapsto 1$}$\fg$,
     $\og${\scriptsize$\Alice \mapsto 2, \Bob \mapsto 0$}$\fg$,\\
  \hspace*{78pt} $\og$ $\scriptstyle \Alice ~\mapsto~ 1, \Bob ~\mapsto~ 2$ $\fg$,  $\fg$.

To describe a specific infinite strategy profile one uses most of the time a fixpoint equation like:
\[t \quad = \quad \og \Alice, \rgt{}, \og {\scriptstyle \Alice ~\mapsto~ 0, \Bob ~\mapsto~ 0}\fg, \og \Bob, \rgt{}, t, t\fg\fg\]
which corresponds to the pictures:
\[
\raisebox{50pt}{
  \begin{psmatrix}[colsep=3pt,rowsep=6pt]
  &&&&[name=0]\\
  &&&&\textit{t}\\
  &[name=1]&&&&&&&[name=2]
  \ncline{0}{1}
  \ncline{0}{2}
  \ncline{1}{2}
\end{psmatrix}
}
\quad \raisebox{75pt}{=} \ \
\begin{psmatrix}[colsep=3pt,rowsep=6pt]
  &&& {\ovalnode{sommet}{A_1}}\\
  & [name=1]{\raisebox{-10pt}{$\scriptscriptstyle A_1 ~\mapsto~ 0, A_2 ~\mapsto~ 0$}} &&&& {\ovalnode{2}{A_2}}\\
  &&&[name=21]&&&&&& [name=22]\\
  &&& t &&&&&&& t\\
  && [name=211] && [name=212] && [name=221] &&&&&&&& [name=222] & 
  \ncline[linewidth=.07]{sommet}{1}
  \ncline{sommet}{2}
  \ncline[linewidth=.07]{2}{21}
  \ncline{sommet}{2}
  \ncline{2}{22}
  \ncline{21}{211}
  \ncline{21}{212}
  \ncline{211}{212}
  \ncline{22}{221}
  \ncline{22}{222}
  \ncline{221}{222}
\end{psmatrix}
\]

Other examples of infinite strategy profiles are given in Section~\ref{sec:dol}.  Usually an infinite game is defined as a cofixpoint, \ie as the solution of an equation, possibly a parametric equation.

Whereas in the finite case one can easily associate with a strategy profile a utility function, \ie a function which assigns a utility to an agent, as
the result of a recursive evaluation, this is no more the case with infinite strategy profiles.  One reason is that it is no more the case that the
utility function can be computed since the strategy profile may run for ever.  This makes the function partial\footnote{Assigning arbitrarily (i.e.,
  not algorithmically) a utility function to an infinite ``history'', as it is made sometimes in the literature, is artificial and not really handy
  for formal reasoning.} and it cannot be defined as an inductive or a coinductive.  Therefore we make $s2u$ (an abbreviation for
\emph{Strategy-profile-to-Utility}) a relation between a strategy profile and a utility function and we define it coinductively; $s2u$ appears in
expression of the form\footnote{Notice the lighter notation $(f~x~y~z)$ for what is usually written $f(x)(y)(z)$.} $(s2u\ s\ a\ u)$ where $s$ is a
strategy profile, $a$ is an agent and $u$ is a utility.  It reads ``$u$ is a utility of the agent $a$ in the strategy profile~$s$''.

\begin{quotation}
$s2u$ is a predicate defined \textbf{inductively} as follows:
\ifElsev \begin{itemize}[$\star$] \else \begin{itemize} \fi
  \item $s2u \og f \fg~a~(f(a))$ holds,
  \item if $s2u~s_l~a~u$ holds then $s2u~\og a',\lft,s_l,s_r\fg~a~u$ holds,
  \item if $s2u~s_r~a~u$ holds then $s2u~\og a',\rgt{},s_l,s_r\fg~a~u$ holds.
  \end{itemize} 
\end{quotation}

This means the utility of $a$ for the leaf strategy profile $\og f\fg$ is $f(a)$, \ie the value delivered by the function $f$ when applied to $a$.
The utility of $a$ for the strategy profile $\og a',\lft,s_l,s_r\fg$ is $u$ if the utility of $a$ for the strategy profile $s_l$ is $u$. In the case
of $s_0$, the first above strategy profile, one has $s2u~s_0~{\Alice}~2$, which means that, for the strategy profile~$s_0$, the utility of {\Alice} is
$2$.  

For a game there are many associated possible strategy profiles, which have a similar structure, but on the other hand there is a function which
returns a game given a strategy profile. 

\section[Equilibria]{Subgame perfect and Nash equilibria}
\label{sec:equi}

\subsection{Convertibility}
\label{sec:conv}

An important binary relation on strategy profiles  is \emph{convertibility}.  We write $\conva$.  the convertibility of agent $a$.

\begin{quotation}
  The relation $\conva$ is defined \textbf{inductively} as follows:
\ifElsev \begin{itemize}[$\star$] \else \begin{itemize} \fi
  \item $\conva$ is reflexive, i.e., for all $s$, $s \conva s$.
  \item If the node has the same agent as the agent in $\conva$ then the choice may change, i.e.,
    \[
    \prooftree s_1 \conva s_1'\qquad s_2 \conva s_2' %
    \justifies \og a, c, s_1, s_2 \fg \ \conva \ \og a, c', s_1', s_2'\fg
    \endprooftree
    \]

  \item If the node does not have the same agent as in $\conva$, then the choice has to be the same:
    \[
    \prooftree 
  s_1 \conva s_1'\qquad s_2 \conva s_2' %
  \justifies \og a', c, s_1, s_2 \fg \ \conva \ \og a', c, s_1', s_2'\fg
  \endprooftree
  \]
\end{itemize}
\end{quotation}

Roughly speaking two strategy profiles are convertible for $a$ if they change only for the choices for $a$.  Since it is defined inductively, this means that
those changes are finitely many.  We feels that this makes sense since an agent can only conceive finitely many issues.

\subsection{Nash equilibria}
\label{sec:NE}

The notion of Nash equilibrium is translated from the notion in textbooks.   The concept of Nash equilibrium is based on a
comparison of utilities; this assumes that an actual utility exists and therefore this requires convertible strategy profiles to \textit{``lead to a leaf''}.
$s$ is a \emph{Nash equilibrium} if the following implication holds:
\begin{quotation}
  \noindent If $s$ \ltl{} and for all agent~$a$ and for all strategy profile~$s'$ which is convertible to $s$, i.e., $s\conva s'$, and which \ltl{}, if $u$ is the
  utility of~$s$ for~$a$ and $u'$ is the utility of $s$' for $a$, then $u' \leut u$.
\end{quotation}
Roughly speaking this means that a Nash equilibrium is a strategy profile in which no agent has interest to change his choice since doing so he cannot get a
better payoff.

\subsection{Subgame Perfect Equilibria}
\label{sec:sgpe}

In order to insure that $s2u$ has a result we define an operator \ltl{} that says that if one follows the choices shown by the strategy profile
one reaches a leaf, \ie one does not go forever.

\begin{quotation}
The predicate \ltl{} is defined \textbf{inductively} as
\ifElsev \begin{itemize}[$\star$] \else \begin{itemize} \fi
  \item the strategy profile $\og f \fg$ \ltl{},
  \item if $s_l$ \ltl{}, then $\og a, \lft, s_l, s_r\fg$ \ltl{},
  \item if $s_r$ \ltl{}, then $\og a, \rgt{}, s_l, s_r\fg$ \ltl{}.
  \end{itemize}
\end{quotation}

This means that a strategy profile, which is itself a leaf, \ltl{} and if the strategy profile is a node, if the choice is \lft{} and if the left strategy subprofile 
\ltl{} then the whole strategy profile \ltl{} and similarly if the choice is \rgt{}.  

If $s$ is a strategy profile that satisfies the predicate \ltl{} then the utility exists and is unique, in other words:
\begin{itemize}
\item[$`(!)$] For all agent $a$ and for all strategy profile $s$, if $s$ \ltl{} then there exists a utility $u$ which ``is a utility of the agent $a$ in the
  strategy profile~$s$''. 
\item[$`(!)$] For all agent $a$ and for all strategy profile $s$, if $s$ \ltl{}, if ``$u$ is a utility of the agent $a$ in the strategy profile~$s$'' and ``$v$ is a
  utility of the agent $a$ in the strategy profile~$s$'' then $u=v$.
\end{itemize}

This means $s2u$ works like a function on strategy profiles which \emph{lead to a leaf}.
We also consider a predicate \altl{} which means that everywhere in the strategy profile, if one follows the choices, one leads to a leaf.  This property is defined everywhere on an
infinite strategy profile and is therefore coinductive.

\begin{quotation}
  The predicate \altl{} is defined \textbf{coinductively} by saying:
\ifElsev \begin{itemize}[$\star$] \else \begin{itemize} \fi
  \item the strategy profile  $\og f \fg$ \altl{},
  \item for all choice $c$, if $\og a, c, s_l, s_r\fg$ \ltl{}, if $s_l$ \altl{}, if $s_r$ \altl{}, then $\og a, c, s_l, s_r\fg$ \altl{}.
  \end{itemize}
\end{quotation}

This says that a strategy profile, which is a leaf, \altl{} and that a composed strategy profile inherits the predicate from its strategy subprofiles provided itself \ltl{}.

Let us consider now \emph{subgame perfect equilibria}, which we write $SGPE$.  $SGPE$ is a property of strategy profiles. It requires the strategy subprofiles to
fulfill coinductively the same property, namely to be a $SGPE$, and to insure that the strategy profile with the best utility for the node agent to be chosen.
Since both the strategy profile and its strategy subprofiles are potentially infinite, it makes sense to define $SGPE$ coinductively.

\pagebreak[2]
\begin{quotation}
  $SGPE$ is defined \textbf{coinductively} as follows:
\ifElsev \begin{itemize}[$\star$] \else \begin{itemize} \fi
  \item $SGPE \og f\fg$,
  \item if $\og a,\lft,s_l,s_r\fg$ \altl{}, if $SGPE(s_l)$ and $SGPE(s_r)$, if $s2u~s_l~a~u$ and $s2u~s_r~a~v$, if $v\leut u$ \\then $SGPE~\og a,\lft,s_l,s_r\fg$,
  \item if $\og a,\rgt{},s_l,s_r\fg$ \altl{}, if $SGPE(s_l)$ and $SGPE(s_r)$, if $s2u~s_l~a~u$ and $s2u~s_r~a~v$, if $u\leut v$ \\then $SGPE~\og a,\rgt{},s_l,s_r\fg$,
  \end{itemize}
\end{quotation}

This means that a strategy profile, which is a leaf, is a subgame perfect equilibrium.  Moreover if the strategy profile is a node, if the strategy profile \altl{}, if it has agent $a$ and choice \lft, if both strategy subprofiles are subgame perfect equilibria and if the utility of the agent $a$ for the right strategy subprofile is less than this for the left strategy subprofile then the whole strategy profile is a subgame perfect equilibrium and vice versa.  If the choice is \rgt{} this works similarly.

Notice that since we require that the utility can be computed not only for the strategy profile, but for the strategy subprofiles and for the strategy subsubprofiles and so on, we require these strategy profiles not only to \textit{``lead to a leaf''} but to \textit{``always lead to a leaf''}.

We define orders (one for each agent $a$) between strategy profiles which we write $\le_a$.
\begin{quotation}
 \noindent $s' \le_{a} s$ iff : 
 If u (respectively u') is the utility for $a$ in $s$ (resp. $s'$), then $u' \leut u$
\end{quotation}

\begin{prop}
$\le_a$ is an order (the proof is straight forward).
\end{prop}

\begin{prop}
  A subgame perfect equilibrium is a Nash equilibrium.
\end{prop}

\begin{proof}
Suppose that $s$ is a strategy profile which is a $SGPE$ and which has to be proved to be is a Nash equilibrium.

Assuming that $s'$ is a  strategy profile such that $s \conva s'$, let us prove  by induction on $s \conva s'$ that $s' \le_{a} s$:
\begin{itemize}
\item Case $s = s'$,  by reflexivity, $s' \le_{a} s$.
\item Case $s = \og x, \lft ,s_l,s_r\fg$ and $s' = \og x, \lft , s'_l, s'_r \fg$ with ${x\neq a}$. $s \conva s'$ and the definition of $\conva$ imply $s_l
  \conva s'_l$ and $s_r \conva s'_r$.   
$s_l$ which is a strategy subprofile of a $SGPE$ is a $SGPE$ as well.  Hence by induction hypothesis, $s'_l \le_a s_l$.

The utility of $s$ (respectively of $s'$) for $a$ is the utility of $s_l$ (respectively of $s_l'$) for $a$, then $s' \le_a s$.
\item The case $s \mathop{=} \og x, \rgt{} ,s_l,s_r\fg$ and $s' \mathop{=} \og x, \rgt{} , s'_l, s'_r \fg$ is similar.
\item Case $s \mathop{=} \og a, \lft, s_l, s_r \fg$ and $s' \mathop{=} \og a, \rgt{}, s'_l, s'_r \fg$, then $s_l \conva s'_l$ and $s_r \conva s'_r$.  Since
$s$ is a $SGPE$, $s_r \le_a s_l$.

Moreover, since $s_r$ is a $SGPE$, by induction hypothesis, $s'_r \le s_r$.
Hence, by transitivity of $\le_a$, $s'_r \le_a s_l$.
But we know that the utility of $s'$ for $a$ is this of $s'_r$ and the utility of $s$ for $a$ is this of~$s_l$, hence $s' \le_a s$.
\item The case  $s \mathop{=} \og a, \rgt{}, s_l, s_r \fg$ and $s' \mathop{=} \og a, \lft, s'_l, s'_r \fg$ is similar.

\end{itemize}
\end{proof}
The above proof is a  presentation of the formal proof written with the help of the proof assistant \Coq.
Notice that it is by induction on $\conva$ which is possible since $\conva$ is inductively defined.   Notice also that $s$ and $s'$ are potentially infinite.

\section[Dollar auction]{Dollar auction games and Nash equilibria}
\label{sec:dol}

The dollar auction has been presented by \citet{Shubik:1971} as the paradigm of escalation, insisting on its paradoxical aspect. It is a sequential
game presented as an auction in which two agents compete to acquire an object of value $v$ ($v>0$) (see \citet[Ex.
3.13]{gintis00:_game_theor_evolv}).  Suppose that both agents bid $\$ 1$ at each turn. If one of them gives up, the other receives the object and both
pay the amount of their bid.\footnote{In a variant, each bidder, when he bids, puts a dollar bill in a hat or in a piggy bank and their is no return
  at the end of the auction.  The last bidder gets the object.}  For instance, if agent \Alice{} stops immediately, she pays nothing and agent \Bob{},
who acquires the object, has a payoff $v$.  In the general turn of the auction, if  \Alice{} abandons, she looses the auction and has a payoff $-n$ and \Bob{} who has already bid $-n$ has a payoff $v-n$.  At the
next turn after \Alice{} decides to continue, bids $\$ 1$ for this and acquires the object due to \Bob{} stopping, \Alice{} has a payoff $v-(n+1)$ and
\Bob{} has a payoff $-n$.  In our formalization we have considered the \emph{dollar auction} up to infinity.  Since we are interested only by the
``asymptotic'' behavior, we can consider the auction after the value of the object has been passed and the payoffs are negative.  The dollar auction
game can be summarized by Fig.~\ref{fig:dol_auct}.  Notice that we assume that {\Alice} starts.  We have recognized three classes of infinite strategy
profiles, indexed by~$n$:

\figdollar

\begin{enumerate}
\item The strategy profile \emph{always give up}, in which both {\Alice} and \Bob{} stop at each turn, in short \textsf{dolAsBs}$_n$.
\item The strategy profile \emph{{\Alice} stops always and \Bob{} continues always}, in short \textsf{dolAsBc}$_n$.
\item The strategy profile \emph{\Alice{} continues always and \Bob{} stops always}, in short \textsf{dolAcBs}$_n$.
\end{enumerate}
The three kinds of strategy profiles are presented in Fig.~\ref{fig:4_strat}.

\figfourstrat

We have shown\footnote{The proofs are typical uses of the \Coq{} \textsf{cofix} tactic. } that the second and third kinds of strategy profiles, in
which one of the agents always stops and the other continues, are subgame perfect equilibria.  For instance, consider the strategy profile
\textsf{dolAsBc}$_n$.  Assume $\mathit{SGPE}(\textsf{dolAsBc}_{n+1})$.  It works as follows: if $\textsf{dolAsBc}_{n+1}$ is a subgame perfect
equilibrium corresponding to the payoff ${-(v+n+1), -(n+1)}$, then
\[\og \Bob, \lft, \textsf{dolAsBc}_{n+1}, \og \Alice "|->" n+1, \Bob "|->" v+n\fg \fg\] 
is again a subgame perfect equilibrium (since $v+n\ge n+1$) and therefore $\textsf{dolAsBc}_{n}$ is a subgame perfect equilibrium, since again $v+n\ge n+1$.\footnote{Since the
  \textsf{cofix} tactic has been used on a strict strategy subprofile, the reasoning is correct.}  We can conclude that for all $n$, $\textsf{dolAsBc}_{n}$ \emph{is a subgame
  perfect equilibrium}.  In other words, we have assumed that $\mathit{SGPE}(\textsf{dolAsBc}_{n})$ is an \emph{invariant} all along the game and that this invariant is preserved
as we proceed backward, through time, into the game.

With the condition $v>1$, we can prove that \textsf{dolAsBs}$_0$ is not a Nash equilibrium, then as a consequence not a subgame perfect equilibrium.
Therefore, the strategy profile that consists in stopping from the beginning and forever is not a Nash equilibrium, this  contradicts what
is said in the
literature \citep*{Shubik:1971,oneill86:_inten_escal_and_dollar_auction,leininger89:_escal_and_coop_in_confl_situat,gintis00:_game_theor_evolv}.


\section{Why  escalation is rational?}
\label{sec:esc}

Many authors agree (see however \citep{halpern01:_subst_ration_backw_induc,stalnaker98:_belief_revis_in_games}) that choosing a subgame perfect equilibrium  is
rational \citep{aumann95}.  Let us show that this can lead to an escalation.  Suppose I am \Alice{} in the middle of the auction, I have two options that are rational: one
option is to stop right away, since I assume that \Bob{} will continue always.  But the second option says that it could be the case that from now on {\Bob} will stop always
(strategy profile $\textsf{dolAcBs}_n$) and I will always continue which is a subgame perfect equilibrium hence rational.  If {\Bob} acts similarly this is the escalation.  So at
each step an agent can stop and be rational, as well as at each step an agent can continue and be rational; both options make perfect sense.  We claim that human agents reason
coinductively unknowingly.  Therefore, for them, escalation is one of their rational options at least if one considers strictly the rules of the dollar auction game, in particular
with no limit on the bankroll.  Many experiences \citep{colman99:_game_theor_and_its_applic} have shown that human are inclined to escalate or at least to go very far in the
auction when playing the dollar auction game.    We propose the following explanation: the finiteness of the game was not explicit for the participants and for them the game was
naturally infinite.  Therefore they adopted a form of reasoning similar to the one we developed here, probably in an intuitive form and they conclude it was equally rational to continue or to
leave according to their feeling on the threat of their opponent, hence their attitude. 
Actually our theoretical work reconciles experiences with logic,\footnote{A logic which includes coinduction.} and human with
rationality. 

\section{Another example: the infinipede}
\label{sec:anot-exampl-infin}

An often studied extensive game is the so-called centipede\footnote{ A centipede has hundred legs, whereas a millipede has thousand. All belong to the
  group of myriapods which means ``ten thousand legs''.} introduced by \cite{rosenthal81:_games_of_perfec_infor_predat} (see also
\cite{binmore87:_model_ration_player,Colman_rat_back_ind,osborne94:_cours_game_theory}). Whereas centipedes are finite extensive games, we have
studied games with infinitely many ``legs'', which we propose to call \emph{infinipedes}. \emph{Infinipedes} are generalization to infinity of
centipedes.  In infinipedes, we have identified only one subgame perfect equilibrium, namely this where both agents abandon at each turn.  This shows
that even in the infinite generalization, agents are rational if they do not start the game and abandon from the beginning.  Hence the paradox
discussed by the authors still remains, namely the agents do not get the somewhat better payoff, they would get if they would be more flexible with
respect to rationality.  The problem for the agents in the infinipede game is that when they start an infinite game, they do not know when to stop.

We notice the specific status of the strategy profile
\textsf{ac} in which all agents continue forever.  Since \textsf{ac} cannot attribute payoffs to the agents, it cannot be compared with any other
strategy profile and lies isolated in its own attractor (in term of equilibrium).  The headlong run \textsf{ac} is somewhat rational despite it does
not deliver any reward.

\section{Conclusion}
\label{sec:concl}


We have shown that coinduction is the right tool to study infinite structures, e.~g., the infinite \emph{dollar auction game}.  This way we get
results which contradict forty years of claims that escalation is irrational.  We can show where the failure comes from, namely from the fact that
authors have extrapolated on infinite structures results obtained on finite ones.  Actually in a strategy profile in which one of the agents threatens
credibly the other to continue in every case, common sense says that the other agent should abandon at each step (taking seriously the threat), this
is a subgame perfect equilibrium.  If the threat to continue is not credible, the other agent may think that his opponent bluffs and will abandon at
every step from now on, hence a rational attitude for him is to continue.  As a matter of fact, coinduction meets common sense.  Since our reasoning
on infinite games proceeds from future to past, we call \emph{backward coinduction} the new method for proving that a given infinite strategy profile
is a subgame perfect equilibrium.  This study has also demonstrated the use of a proof assistant in such a development. Indeed the results on infinite
objects are sometime so counter-intuitive that a check on a proof assistant is essential.  We think that this opens new perspectives to game theory
toward a more formal approach based on the last advances in mathematics offered by proof assistants
\citep{harrison-notices,dowek07:_les_metam_du_calcul}.  Among others, an issue is to extend  Aumann's connection \citep{aumann95} between subgame perfect equilibria (or
backward coinduction) and coinductively defined common knowledge  \citep{capretta:2007}.


\appendix

\section{Excerpts of Coq scripts}
\label{sec:coq_vern}

\subsubsection*{Infinite binary trees}~\medskip

\noindent
\coqdockw{CoInductive} \coqdocid{LBintree} : \coqdocid{Set} :=\coqdoceol
\noindent
| \coqdocid{LbtNil}: \coqdocid{LBintree} \coqdoceol
\noindent
| \coqdocid{LbtNode}: \coqdocid{LBintree} \ensuremath{\rightarrow} \coqdocid{LBintree} \ensuremath{\rightarrow} \coqdocid{LBintree}.\coqdoceol

\medskip
\noindent
\coqdockw{CoInductive} \coqdocid{InfiniteLBT}: \coqdocid{LBintree} \ensuremath{\rightarrow} \coqdocid{Prop} := \coqdoceol
\noindent
| \coqdocid{IBTLeft} : \ensuremath{\forall} \coqdocid{bl} \coqdocid{br}, \coqdocid{InfiniteLBT} \coqdocid{bl} \ensuremath{\rightarrow} \coqdocid{InfiniteLBT} (\coqdocid{LbtNode} \coqdocid{bl} \coqdocid{br})\coqdoceol
\noindent
| \coqdocid{IBTRight} : \ensuremath{\forall} \coqdocid{bl} \coqdocid{br}, \coqdocid{InfiniteLBT} \coqdocid{br} \ensuremath{\rightarrow} \coqdocid{InfiniteLBT} (\coqdocid{LbtNode} \coqdocid{bl} \coqdocid{br}).\coqdoceol

\medskip
\noindent
\coqdockw{CoFixpoint} \coqdocid{Zig}: \coqdocid{LBintree} := \coqdocid{LbtNode} \coqdocid{Zag} \coqdocid{LbtNil}\coqdoceol
\noindent
\coqdocid{with} \coqdocid{Zag}: \coqdocid{LBintree} := \coqdocid{LbtNode} \coqdocid{LbtNil} \coqdocid{Zig}.\coqdoceol

\subsubsection*{Infinite strategy profiles}~\medskip

\noindent
\coqdockw{CoInductive}
 \coqdocid{StratProf} : \coqdocid{Set} :=\coqdoceol
\noindent
| \coqdocid{sLeaf} : \coqdocid{Utility\_fun} \ensuremath{\rightarrow} \coqdocid{StratProf}\coqdoceol
\noindent
| \coqdocid{sNode} : \coqdocid{Agent} \ensuremath{\rightarrow} \coqdocid{Choice} \ensuremath{\rightarrow} \coqdocid{StratProf} \ensuremath{\rightarrow} \coqdocid{StratProf} \ensuremath{\rightarrow} \coqdocid{StratProf}.\coqdoceol

\medskip

\noindent
\coqdockw{Inductive} \coqdocid{s2u} : \coqdocid{StratProf} \ensuremath{\rightarrow} \coqdocid{Agent} \ensuremath{\rightarrow} \coqdocid{Utility} \ensuremath{\rightarrow} \coqdocid{Prop} :=\coqdoceol
\noindent
| \coqdocid{s2uLeaf}: \ensuremath{\forall} \coqdocid{a} \coqdocid{f}, \coqdocid{s2u} ($\og$ \coqdocid{f}$\fg$) \coqdocid{a} (\coqdocid{f} \coqdocid{a})\coqdoceol
\noindent
| \coqdocid{s2uLeft}: \ensuremath{\forall}  (\coqdocid{a} \coqdocid{a'}:\coqdocid{Agent}) (\coqdocid{u}:\coqdocid{Utility}) (\coqdocid{sl} \coqdocid{sr}:\coqdocid{StratProf}),\coqdoceol
\coqdocindent{2.00em}
\coqdocid{s2u} \coqdocid{sl} \coqdocid{a} \coqdocid{u}  \ensuremath{\rightarrow} \coqdocid{s2u} ($\og$ \coqdocid{a'},\coqdocid{l},\coqdocid{sl},\coqdocid{sr}$\fg$) \coqdocid{a} \coqdocid{u}  \coqdoceol
\noindent
| \coqdocid{s2uRight}: \ensuremath{\forall} (\coqdocid{a} \coqdocid{a'}:\coqdocid{Agent}) (\coqdocid{u}:\coqdocid{Utility}) (\coqdocid{sl} \coqdocid{sr}:\coqdocid{StratProf}),\coqdoceol
\coqdocindent{2.00em}
\coqdocid{s2u} \coqdocid{sr} \coqdocid{a} \coqdocid{u} \ensuremath{\rightarrow} \coqdocid{s2u} ($\og$ \coqdocid{a'},\coqdocid{r},\coqdocid{sl},\coqdocid{sr}$\fg$) \coqdocid{a} \coqdocid{u}.\coqdoceol

\medskip
\noindent
\coqdockw{Inductive} \coqdocid{LeadsToLeaf}: \coqdocid{StratProf} \ensuremath{\rightarrow} \coqdocid{Prop} :=\coqdoceol
\noindent
| \coqdocid{LtLLeaf}: \ensuremath{\forall} \coqdocid{f}, \coqdocid{LeadsToLeaf} ($\og$ \coqdocid{f}$\fg$)\coqdoceol
\noindent
| \coqdocid{LtLLeft}: \ensuremath{\forall} (\coqdocid{a}:\coqdocid{Agent})(\coqdocid{sl}: \coqdocid{StratProf}) (\coqdocid{sr}:\coqdocid{StratProf}),\coqdoceol
\coqdocindent{2.00em}
\coqdocid{LeadsToLeaf} \coqdocid{sl} \ensuremath{\rightarrow} \coqdocid{LeadsToLeaf} ($\og$ \coqdocid{a},\coqdocid{l},\coqdocid{sl},\coqdocid{sr}$\fg$)\coqdoceol
\noindent
| \coqdocid{LtLRight}: \ensuremath{\forall} (\coqdocid{a}:\coqdocid{Agent})(\coqdocid{sl}: \coqdocid{StratProf}) (\coqdocid{sr}:\coqdocid{StratProf}),\coqdoceol
\coqdocindent{2.50em}
\coqdocid{LeadsToLeaf} \coqdocid{sr} \ensuremath{\rightarrow} \coqdocid{LeadsToLeaf} ($\og$ \coqdocid{a},\coqdocid{r},\coqdocid{sl},\coqdocid{sr}$\fg$).\coqdoceol

\medskip
\noindent
\coqdockw{CoInductive} \coqdocid{AlwLeadsToLeaf}: \coqdocid{StratProf} \ensuremath{\rightarrow} \coqdocid{Prop} :=\coqdoceol
\noindent
| \coqdocid{ALtLeaf} : \ensuremath{\forall} (\coqdocid{f}:\coqdocid{Utility\_fun}), \coqdocid{AlwLeadsToLeaf} ($\og$\coqdocid{f}$\fg$)\coqdoceol
\noindent
| \coqdocid{ALtL} : \ensuremath{\forall} (\coqdocid{a}:\coqdocid{Agent})(\coqdocid{c}:\coqdocid{Choice})(\coqdocid{sl} \coqdocid{sr}:\coqdocid{StratProf}),\coqdoceol
\coqdocindent{2.00em}
\coqdocid{LeadsToLeaf} ($\og$\coqdocid{a},\coqdocid{c},\coqdocid{sl},\coqdocid{sr}$\fg$) \ensuremath{\rightarrow} \coqdocid{AlwLeadsToLeaf} \coqdocid{sl} \ensuremath{\rightarrow}\coqdocid{AlwLeadsToLeaf} \coqdocid{sr} \ensuremath{\rightarrow} \coqdoceol
\coqdocindent{2.00em}
\coqdocid{AlwLeadsToLeaf} ($\og$\coqdocid{a},\coqdocid{c},\coqdocid{sl},\coqdocid{sr}$\fg$).\coqdoceol

\subsubsection*{SGPE}~\medskip

  \noindent \coqdockw{CoInductive} \coqdocid{SGPE}: \coqdocid{StratProf} \ensuremath{\rightarrow} \coqdocid{Prop} :=\coqdoceol
  \noindent | \coqdocid{SGPE\_leaf}: \ensuremath{\forall} \coqdocid{f}:\coqdocid{Utility\_fun}, \coqdocid{SGPE} ($\og$\coqdocid{f}$\fg$)\coqdoceol
  \noindent | \coqdocid{SGPE\_left}: \ensuremath{\forall} (\coqdocid{a}:\coqdocid{Agent})(\coqdocid{u} \coqdocid{v}: \coqdocid{Utility}) (\coqdocid{sl} \coqdocid{sr}:
  \coqdocid{StratProf}), \coqdoceol \coqdocindent{2.00em} \coqdocid{AlwLeadsToLeaf} ($\og$\coqdocid{a},\coqdocid{l},\coqdocid{sl},\coqdocid{sr}$\fg$) \ensuremath{\rightarrow}
  \coqdoceol \coqdocindent{2.00em} \coqdocid{SGPE} \coqdocid{sl} \ensuremath{\rightarrow} \coqdocid{SGPE} \coqdocid{sr} \ensuremath{\rightarrow} \coqdoceol \coqdocindent{2.00em}
  \coqdocid{s2u} \coqdocid{sl} \coqdocid{a} \coqdocid{u} \ensuremath{\rightarrow} \coqdocid{s2u} \coqdocid{sr} \coqdocid{a} \coqdocid{v} \ensuremath{\rightarrow} (\coqdocid{v}
  \leut \coqdocid{u}) \ensuremath{\rightarrow} \coqdoceol \coqdocindent{2.00em} \coqdocid{SGPE} ($\og$\coqdocid{a},\coqdocid{l},\coqdocid{sl},\coqdocid{sr}$\fg$)\coqdoceol
  \noindent | \coqdocid{SGPE\_right}: \ensuremath{\forall} (\coqdocid{a}:\coqdocid{Agent}) (\coqdocid{u} \coqdocid{v}:\coqdocid{Utility}) (\coqdocid{sl} \coqdocid{sr}:
  \coqdocid{StratProf}), \coqdoceol \coqdocindent{2.00em} \coqdocid{AlwLeadsToLeaf} ($\og$\coqdocid{a},\coqdocid{r},\coqdocid{sl},\coqdocid{sr}$\fg$) \ensuremath{\rightarrow}
  \coqdoceol \coqdocindent{2.00em} \coqdocid{SGPE} \coqdocid{sl} \ensuremath{\rightarrow} \coqdocid{SGPE} \coqdocid{sr} \ensuremath{\rightarrow} \coqdoceol \coqdocindent{2.00em}
  \coqdocid{s2u} \coqdocid{sl} \coqdocid{a} \coqdocid{u} \ensuremath{\rightarrow} \coqdocid{s2u} \coqdocid{sr} \coqdocid{a} \coqdocid{v} \ensuremath{\rightarrow} (\coqdocid{u}
  \leut \coqdocid{v}) \ensuremath{\rightarrow} \coqdoceol \coqdocindent{2.00em} \coqdocid{SGPE} ($\og$\coqdocid{a},\coqdocid{r},\coqdocid{sl},\coqdocid{sr}$\fg$). \coqdoceol

\subsubsection*{Nash equilibrium}~\medskip

\noindent
\coqdockw{Definition} \coqdocid{NashEq} (\coqdocid{s}: \coqdocid{StratProf}): \coqdocid{Prop} := \coqdoceol
\coqdocindent{1.00em}
\ensuremath{\forall} \coqdocid{a} \coqdocid{s'} \coqdocid{u} \coqdocid{u'}, \coqdocid{s'}\conva\coqdocid{s} \ensuremath{\rightarrow} \coqdoceol
\coqdocindent{2.00em}
\coqdocid{LeadsToLeaf} \coqdocid{s'} \ensuremath{\rightarrow} (\coqdocid{s2u} \coqdocid{s'} \coqdocid{a} \coqdocid{u'}) \ensuremath{\rightarrow} \coqdoceol
\coqdocindent{2.00em}
\coqdocid{LeadsToLeaf} \coqdocid{s} \ensuremath{\rightarrow} (\coqdocid{s2u} \coqdocid{s} \coqdocid{a} \coqdocid{u}) \ensuremath{\rightarrow} \coqdoceol
\coqdocindent{2.00em}
(\coqdocid{u'} \leut \coqdocid{u}).\coqdoceol

\subsubsection*{{\Alice} stops always and \Bob{} continues always}~\medskip

\noindent
\coqdockw{Definition} \coqdocid{add\_Alice\_Bob\_dol} (\coqdocid{cA} \coqdocid{cB}:\coqdocid{Choice}) (\coqdocid{n}:\coqdocid{nat}) (\coqdocid{s}:\coqdocid{Strat}) :=\coqdoceol
\coqdocindent{1.00em}
\og\coqdocid{Alice},\coqdocid{cA},\og\coqdocid{Bob}, \coqdocid{cB},\coqdocid{s},[\coqdocid{n}+1, \coqdocid{v}+\coqdocid{n}]\fg,[\coqdocid{v}+\coqdocid{n},\coqdocid{n}]\fg.\coqdoceol

\medskip
\noindent
\coqdockw{CoFixpoint} \coqdocid{dolAcBs} (\coqdocid{n}:\coqdocid{nat}): \coqdocid{Strat} := \coqdocid{add\_Alice\_Bob\_dol} \coqdocid{l} \coqdocid{r} \coqdocid{n} (\coqdocid{dolAcBs} (\coqdocid{n}+1)).\coqdoceol

\medskip
\noindent
\coqdockw{Theorem} \coqdocid{SGPE\_dol\_Ac\_Bs}:  \ensuremath{\forall} (\coqdocid{n}:\coqdocid{nat}), \coqdocid{SGPE} \coqdocid{ge} (\coqdocid{dolAcBs} \coqdocid{n}).\coqdoceol

\subsubsection*{\Alice{} continues always and \Bob{} stops always}~\medskip

\noindent
\coqdockw{CoFixpoint} \coqdocid{dolAsBc} (\coqdocid{n}:\coqdocid{nat}): \coqdocid{Strat} := \coqdocid{add\_Alice\_Bob\_dol} \coqdocid{r} \coqdocid{l} \coqdocid{n} (\coqdocid{dolAsBc} (\coqdocid{n}+1)).\coqdoceol

\medskip
\noindent
\coqdockw{Theorem} \coqdocid{SGPE\_dol\_As\_Bc}:  \ensuremath{\forall} (\coqdocid{n}:\coqdocid{nat}), \coqdocid{SGPE} \coqdocid{ge} (\coqdocid{dolAsBc} \coqdocid{n}).\coqdoceol

\subsubsection*{Always give up}~\medskip

\noindent
\coqdockw{CoFixpoint} \coqdocid{dolAsBs} (\coqdocid{n}:\coqdocid{nat}): \coqdocid{Strat} := \coqdocid{add\_Alice\_Bob\_dol} \coqdocid{r} \coqdocid{r} \coqdocid{n} (\coqdocid{dolAsBs} (\coqdocid{n}+1)).\coqdoceol

\medskip
\noindent
\coqdockw{Theorem} \coqdocid{NotSGPE\_dolAsBs}: (\coqdocid{v}$>$1) \ensuremath{\rightarrow} \~{}(\coqdocid{NashEq} \coqdocid{ge} (\coqdocid{dolAsBs} 0)).\coqdoceol

\end{document}